\documentclass[aps,twocolumn,showpacs,superscriptaddress]{revtex4}

\usepackage{graphicx}
\usepackage{bm}
\usepackage{amsmath}
\usepackage{dcolumn}%

\newcolumntype{/}{D{/}{/}{2,2}}  
\newcolumntype{.}{D{.}{.}{0}}  

\begin{document}

\title{Resonant inelastic x-ray scattering spectra in the hyperhoneycomb
  iridate $\beta$-Li$_2$IrO$_3$: First principles calculations}

\author{V.N. Antonov}

\affiliation{G. V. Kurdyumov Institute for Metal Physics of the
  N.A.S. of Ukraine, 36 Academician Vernadsky Boulevard, UA-03142
  Kyiv, Ukraine}

\affiliation{Faculty of Physics, University of Bialystok,
  K. Ciolkowskiego 1L, PL-15-245 Bialystok, Poland}

\author{D.A. Kukusta}

\affiliation{G. V. Kurdyumov Institute for Metal Physics of the
  N.A.S. of Ukraine, 36 Academician Vernadsky Boulevard, UA-03142
  Kyiv, Ukraine}

\affiliation{Max-Planck-Institut f\"ur Festk\"orperforschung,
Heisenberg Strasse 1, D-70569 Stuttgart, Germany}

\author{L. Uba} 

\affiliation{Faculty of Physics, University of Bialystok,
  K. Ciolkowskiego 1L, PL-15-245 Bialystok, Poland}

\author{A. Bonda} 

\affiliation{Faculty of Physics, University of Bialystok,
  K. Ciolkowskiego 1L, PL-15-245 Bialystok, Poland}

\author{S. Uba} 

\affiliation{Faculty of Physics, University of Bialystok,
  K. Ciolkowskiego 1L, PL-15-245 Bialystok, Poland}

\date{\today}

\begin{abstract}

We studied the electronic structure of $\beta$-Li$_2$IrO$_3$ insulator
within the density-functional theory using the generalized gradient
approximation with taking into account strong Coulomb correlations in
the framework of the fully relativistic spin-polarized Dirac linear
muffin-tin orbital band-structure method. The $\beta$-Li$_2$IrO$_3$
undergoes a pressure-induced structural and magnetic phase transitions
at $P_c$ $\sim$4 GPa with symmetry lowering to the monoclinic
$C2/c$. The structural phase transition is accompanied by the
formation of Ir$_2$ dimers on the zigzag chains, with an Ir-Ir
distance of $\sim$2.66~\AA, even shorter than that of metallic Ir. The
strong dimerization stabilizes the bonding molecular-orbital state,
leads to the collapse of the magnetism and opens the energy gap with a
concomitant electronic phase transition from a Mott insulator to band
insulator. The resonant inelastic x-ray scattering spectra (RIXS) at
the Ir $L_3$ edge were investigated theoretically from first
principles. The calculated results are in good agreement with the
experimental data. We show that the the drastic reconstruction of the
RIXS spectral peak at 0.7 eV associated with the structural $Fddd
\rightarrow C2/c$ phase transition at $P_c$ can be related to
disappearing of the Coulomb correlations in the high-pressure $C2/c$
phase.

\end{abstract}


\maketitle

\section{Introduction}

\label{sec:introd}

Quantum spin liquids (QSLs) \cite{JaKh09,WCK+14,KAV14} represent a
novel state of matter in which quantum fluctuations prevent the
conventional magnetic order from being established, and the spins
remain disordered even at zero temperature. It is an emerging and fast
growing field. In this context the Kitaev Hamiltonian (KH) on
honeycomb lattices has great promise \cite{Kit06}. The paramount
attention given to such states can be understood by the fact that they
may be topologically protected from decoherence \cite{AHM+16}, display
fractional excitations with Majorana statistics, and therefore hold
promise in the field of quantum information and quantum computation
\cite{WTD+17,SLY+16}. The field of QSLs is still wide open, both
theoretically and experimentally. So far, there have been a new
experimental discoveries and theoretical ideas are rapidly
emerging. However, a basic mathematical framework that can be used to
understand QSLs systematically is still lacking. The major difficulty
in understanding QSLs is that they are intrinsically strongly
correlated systems, for which no perturbative approach is available.

Possible realization of such an exotic state has been suggested in
Mott insulators such as 213 iridates A$_2$IrO$_3$ (with A = Na, Li),
which have a honeycomb layered structure consisting of IrO$_6$
octahedra \cite{YCC+12,WTD+17}. They have drawn much attention as a
candidate for topological insulators \cite{SKK+09,HaKa10} with
electron correlations. Both nontrivial hopping terms induced by the
strong spin orbit (SO) coupling and significant on-site Coulomb
correlations make honeycomb iridates a possible candidate also for the
compounds with Kitaev spin liquid type ground state
\cite{CJK13,CQL+13,BJC+14,KCK+14,TKD+15,SLLK15,GLD+16}. It was
proposed that strong SO interaction in these iridates reorganizes the
crystal field states of the 5$d$ orbitals into a $J$-multiplet
structure, where $J$ is the combined spin and effective orbital
angular momentum. In this case, the Ir $t_{2g}$ bands are most
naturally described by relativistic atomic orbitals with the effective
total angular momentum, $J_\text{eff}$=3/2 and $J_\text{eff}$=1/2. In
this approximation, the splitting between the 3/2 and 1/2 states is
larger than their dispersion. The $J_\text{eff}$=1/2 band is
half-filled and the relatively weak Coulomb repulsion $U$ is
sufficient to split the $J_{eff}$ = 1/2 doublet into lower and upper
Hubbard bands, giving rise to a novel Mott insulator \cite{KJM+08}.

There are several recent publications on the experimental and
theoretical investigations of the electronic structure and various
physical properties of honeycomb iridate Li$_2$IrO$_3$
\cite{KKK16,BJC+14,LeKi15,KCV15,KYH+16,RFB+17,DRP18,MMS+18,HAE+18}.
However, the long-sought spin-liquid state has remained elusive. All
three honeycomb polytypes of Li$_2$IrO$_3$ (including $\alpha$,
$\beta$, and $\gamma$ phases) are magnetically ordered that suggests
that the Heisenberg interaction is still sizable. In addition,
trigonal crystal fields also compete with the Kitaev interaction.

In the present study, we focus our attention on the theoretical
investigation of the resonant inelastic x-ray scattering (RIXS)
spectra in the $\beta$-Li$_2$IrO$_3$ compound from first
principles. RIXS is a fast developing experimental technique in which
one scatters x-ray photons inelastically off matter. It is a photon-in
photon-out spectroscopy for which one can, in principle, measure the
energy, momentum, and polarization change of the scattered
photon. Compared to other scattering techniques, RIXS has number of
unique features. It covers a large scattering phase space, is
polarization dependent, element and orbital specific, bulk sensitive
\cite{AVD+11}. The RIXS spectra at the Ir $L3$ edge in
$\beta$-Li$_2$IrO$_3$ were measured by Takayama {\it et al.}
\cite{TKG+19} as a function of applied hydrostatic pressure. The
$\beta$-Li$_2$IrO$_3$ undergoes a pressure-induced structural phase
transition at $P_c$ $\sim$4 GPa with symmetry lowering to the
monoclinic $C2/c$. The structural phase transition is accompanied by
the formation of Ir$_2$ dimers on the zigzag chains, with an Ir-Ir
distance of $\sim$2.66 \AA, even shorter than that of metallic Ir. The
strong dimerization stabilizes the bonding molecular-orbital state,
leads to the collapse of the magnetism and opens the energy gap with a
concomitant electronic phase transition from a Mott insulator to band
insulator \cite{TKG+19}. The experimental measurements showed the
drastic reconstruction of the RIXS spectra associated with this
dimerization \cite{TKG+19}. With increasing pressure above $P_c$ the
prominent peak at $\sim$0.7 eV is suppressed strongly. The one at 3.5
eV is broadened but remains in the high-pressure phase. There are some
changes at 0.5 to 2.0 eV energy interval, and a shoulder-like feature
around 2.8 eV emerges. The aim of the present work is to investigate
the RIXS spectra in $\beta$-Li$_2$IrO$_3$ compound from the first
principles and its evolution under pressure-induced structural phase
transitions.

The paper is organized as follows. The computational details are
presented in Sec. II. Sec. III presents the electronic structure and
theoretically calculated RIXS spectra of the $\beta$-Li$_2$IrO$_3$
compound compared with the experimental measurements. Finally, the
results are summarized in Sec. IV.

\section{Computational details}
\label{sec:details}

\paragraph{Crystal structure.} 

The $\beta$-Li$_2$IrO$_3$ crystallizes in the orthorhombic space group
$Fddd$, with zigzag chains running in alternating directions (see
Fig. 1 in Ref. \cite{AUU18} as well as Fig. 1 in
Ref. \cite{TKG+19}). In the language of the Kitaev interactions, these
chains form the $x$- and $y$-bonds, while the $z$-bonds link together
adjacent layers of chains. In the hyperhoneycomb Ir sublattice of
$\beta$-Li$_2$IrO$_3$, the zigzag Ir chains are connected by the
bridging bonds parallel to the $c$ axis, all the angles between the
three Ir-Ir bonds are close to 120$^{\circ}$, and the distances
between Ir atoms are almost equal (only $\sim$0.2\% difference). At
ambient pressure, Ir-Ir bonds along zig-zag chains have a length
$d_{x,y}$ =2.9729 \AA, and bonds between the chains $d_{z}$ =2.9784
\AA.

As pressure increases from ambient pressure to 3.08 GPa, the $d_{x,y}$
Ir-Ir bonds shrink from 2.9729 to 2.9246 \AA; the corresponding
$d_{z}$ Ir-Ir bonds also decrease from 2.9784 to 2.9379 \AA\,
\cite{VEG+17,TKG+19}. At $P_c$, the bonds between the chains $d_{z}$
slightly increase to 3.0129 \AA; one of the $x/y$ bonds also increases
to 3.0143 \AA, while the other one decreases strongly to 2.6609
\AA. Note that this distance is even smaller than the Ir-Ir distance
of 2.714 \AA\, in Ir metal. Such a remarkably small interatomic
distance strongly suggests the formation of Ir$_2$ dimers at $P_c$
\cite{TKG+19}. The pressure dependence of the structural parameters of
the orthorhombic $\beta$-Li$_2$IrO$_3$ can be found in
Refs. \cite{VEG+17,TKG+19}.

\paragraph{Resonant inelastic x-ray scattering.} 

In the direct RIXS process \cite{AVD+11} an incoming photon with
energy $\hbar \omega_{\mathbf{k}}$, momentum $\hbar \mathbf{k}$ and
polarization $\bm{\epsilon}$ excites the solid from a ground state
$|{\rm g}\rangle$ with energy $E_{\rm g}$ to the intermediate state
$|{\rm I}\rangle$ with energy $E_{\rm I}$. During relaxation the
outcoming photon with energy $\hbar \omega_{\mathbf{k}'}$, momentum
$\hbar \mathbf{k}'$ and polarization $\bm{\epsilon}'$ is emitted, and
the solid is in the state $|{\rm f}\rangle$ with energy $E_{\rm
  f}$. As a result an excitation with energy $\hbar \omega = \hbar
\omega_{\mathbf{k}} - \hbar \omega_{\mathbf{k}'}$ and momentum $\hbar
\mathbf{q}$ = $\hbar \mathbf{k} - \hbar \mathbf{k}'$ is created.  Our
implementation of the code for calculation of the RIXS intensity uses
Dirac four-component basis functions \cite{NKA+83} in the perturbative
approach \cite{ASG97}. RIXS is the second-order process, and its
intensity is given by

\begin{eqnarray}
I(\omega, \mathbf{k}, \mathbf{k}', \bm{\epsilon}, \bm{\epsilon}')
&\propto&\sum_{\rm f}\left| \sum_{\rm I}{\langle{\rm
    f}|\hat{H}'_{\mathbf{k}'\bm{\epsilon}'}|{\rm I}\rangle \langle{\rm
    I}|\hat{H}'_{\mathbf{k}\bm{\epsilon}}|{\rm g}\rangle\over
  E_{\rm g}-E_{\rm I}} \right|^2 \nonumber \\ && \times
\delta(E_{\rm f}-E_{\rm g}-\hbar\omega),
\label{I1}
\end{eqnarray}
where the delta function enforces energy conservation, and the RIXS
perturbation operator in the dipolar approximation is given by the
lattice sum $\hat{H}'_{\mathbf{k}\bm{\epsilon}}=
\sum_\mathbf{R}\hat{\bm{\alpha}}\bm{\epsilon} \exp(-{\rm
  i}\mathbf{k}\mathbf{R})$, where $\hat{\bm{\alpha}}$ are Dirac
matrices. Both $|{\rm g}\rangle$ and $|{\rm f}\rangle$ states are
dispersive so the sum over final states is calculated using the linear
tetrahedron method \cite{LeTa72}.

Detailed expressions for the matrix elements in the electric dipole
approximation in the framework of fully relativistic Dirac
representation were presented in Ref. \cite{unpub:KuYar18}.

\paragraph{Calculation details}

The details of the computational method are described in our previous
papers \cite{AJY+06,AHY+07b,AYJ10,AKB20} and here we only mention
several aspects. Band structure calculations were performed using the
fully relativistic linear muffin-tin orbital (LMTO) method
\cite{And75,PYLMTO}. This implementation of the LMTO method uses
four-component basis functions constructed by solving the Dirac
equation inside an atomic sphere \cite{NKA+83}.  The
exchange-correlation functional of a GGA-type was used in the version
of Perdew, Burke and Ernzerhof (PBE) \cite{PBE96}. Brillouin zone (BZ)
integration was performed using the improved tetrahedron method
\cite{BJA94}. The basis consisted of Ir $s$, $p$, $d$, and $f$; and Li
and O $s$, $p$, and $d$ LMTO's.

To take into account electron-electron correlation effects, we used in
this work the "relativistic" generalization of the rotationally
invariant version of the LSDA+$U$ method \cite{YAF03} which takes into
account SO coupling so that the occupation matrix of localized
electrons becomes non-diagonal in spin indexes. We use in our
calculations the value of $U_{\text{eff}}$ = 1.5 eV ($U$=2.15 eV and
$J_H$=0.65 eV) which gives the best agreement between the calculated
and experimental optical spectra in the $\beta$-Li$_2$IrO$_3$
\cite{AUU18}.

We used in our calculations vector {\bf q} = (0, 10, 0). Takayama {\it
  et al.}  \cite{TKG+19} show that at a low pressure of 0.9 GPa, the
RIXS spectrum at the Ir $L3$ for the single crystal agrees well with
that of the polycrystalline sample at ambient pressure. This supports
the idea that the $d-d$ excitations show only a small {\bf q}
dependence in $\beta$-Li$_2$IrO$_3$.

\section{Electronic structure}
\label{sec:bands}

\begin{figure}[tbp!]
\begin{center}
\includegraphics[width=0.99\columnwidth]{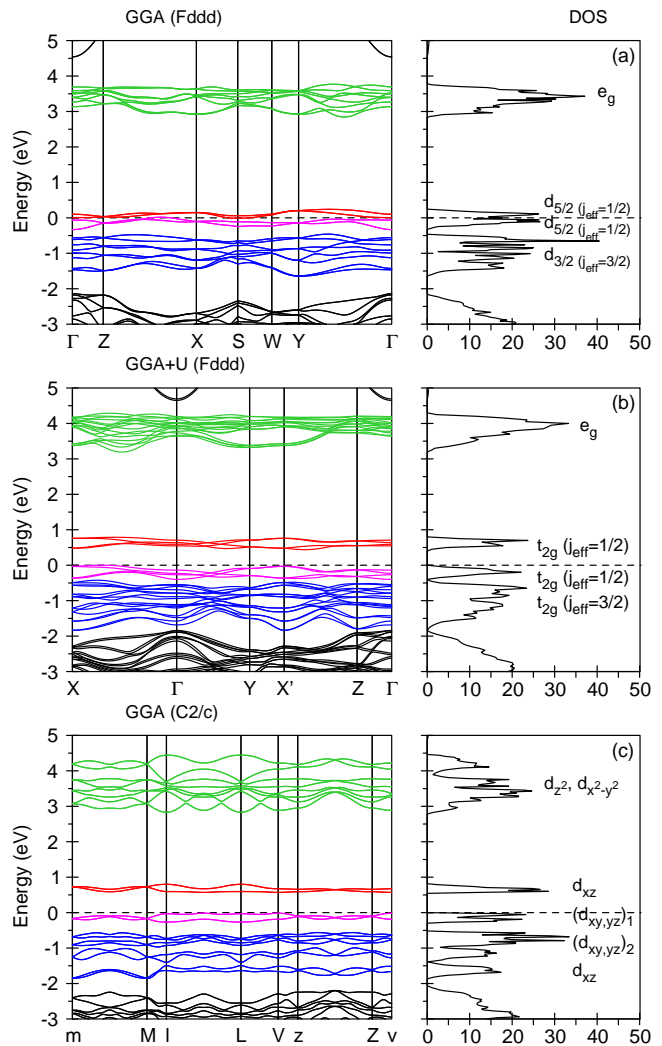}
\end{center}
\caption{\label{BND}(Color online) (a) The {\it ab initio} energy band
  structure of the $\beta$-Li$_2$IrO$_3$ for the ambient pressure
  phase ($Fddd$) in the fully relativistic Dirac GGA+SO approximation;
  (b) the energy band structure of the $\beta$-Li$_2$IrO$_3$ for the
  ambient pressure phase ($Fddd$) in the GGA+SO+$U$ approximation; (c)
  the energy band structure of the $\beta$-Li$_2$IrO$_3$ for the
  high-pressure phase ($C2/c$) in the GGA+SO approach.  }
\end{figure}

Figure \ref{BND} presents the {\it ab initio} energy band structure of
the $\beta$-Li$_2$IrO$_3$ in the energy range of $-$3 to 5 eV for the
ambient pressure $Fddd$ phase, calculated in the fully relativistic
Dirac GGA+SO approximation (the upper panel) and with taking into
account Coulomb correlations in the GGA+SO+$U$ approximation (middle
panel). The lower panel presents the energy band structure of the
$\beta$-Li$_2$IrO$_3$ for the high-pressure phase ($C2/c$) in the
GGA+SO approach.

In Li$_2$IrO$_3$ each Ir$^{4+}$ ion surrounded by six O$^{2-}$ ions
has five valent 5$d$ electrons. The octahedral crystal field largely
splits the Ir $t_{2g}$ and $e_g$ manifolds, so that all five electrons
occupy the $t_{2g}$ manifold. As a result of strong SO coupling
($\Delta_{SO}\sim$ 0.78 eV), the six $t_{2g}$ orbitals are further
separated into two manifolds with $J_\text{eff}$ = 3/2 and
$J_\text{eff}$ = 1/2 [Fig. \ref{BND}(a)]. The $J_\text{eff}$ = 3/2
states are fully filled and the $J_\text{eff}$ = 1/2 states are half
filled. The functions of the $J_\text{eff}$ = 3/2 quartet are
dominated by $d_{3/2}$ states with some minor influence of $d_{5/2}$
states, which is determined by the relative strengths of SO coupling
and crystal-field splitting. The $J_\text{eff}$ = 1/2 functions, on
the other hand, are given by linear combinations of $d_{5/2}$ states
only. This allows one to identify bands with pure $d_{5/2}$ character
as originating from $J_\text{eff}$ = 1/2 states.

The GGA+SO approximation produces a metallic ground state in the
$\beta$-Li$_2$IrO$_3$ [Fig. \ref{BND}(a)], in contradiction with
resistivity measurements which claim that the $\beta$-Li$_2$IrO$_3$ is
a Mott insulator. To produce the correct ground state we have to take
into account Hubbard electron-electron correlations. The
$J_{\text{eff}}$ = 1/2 spin-orbit integrated states form a narrow band
so that even small $U_\text{eff}$=1.5 eV opens Mott gap and splits the
5$d_{5/2}$ ($J_\text{eff}$ = 1/2) band into the upper Hubbard band
(UHB) above the Fermi level [red lines in Fig. \ref{BND}(b)] and lower
Hubbard band (LHB) below the Fermi level [magenta lines in
  Fig. \ref{BND}(b)]. The formation of the $J_\text{eff}$ bands is a
natural consequence of the $J_\text{eff}$ Hubbard model. The $e_g$
orbitals are almost degenerate at ambient pressure occupying the
3.4$-$4.2 eV energy interval.

The $\beta$-Li$_2$IrO$_3$ compound shows strong SO coupling concurrent
with electronic correlations, leading to interesting electronic and
magnetic properties. Due to a delicate balance between these
interactions, minor changes in its parameters may result in a drastic
alteration of the magnetic ground state. These changes may deviate
from long-range magnetic order and perhaps lead towards unexplored
phases such as a quantum spin liquid. One of the experimental
approaches suited to achieving this goal is the application of an
external hydrostatic pressure. The application of external pressure is
a very efficient and clean way to adjust the ground state of materials
without introducing additional scattering centers. Hydrostatic
pressure was applied to the hyperhoneycomb material,
$\beta$-Li$_2$IrO$_3$, and it was found that the crystal structure of
the $\beta$-Li$_2$IrO$_3$ is transformed from the orthorhombic $Fddd$
symmetry to the monoclinic $C2/c$ one at $P_c$ $\sim$4 GPa
\cite{VEG+17,TKG+19}. Pressure reduces the tendency toward magnetism,
thus diminishing the energetic advantage of forming an
antiferromagnetic state, and it brings Ir ions closer together,
enhancing the advantage of forming covalent bonds. The structural
$Fddd \rightarrow C2/c$ phase transition at $P_c$ $\sim$4 GPa is
accompanied by a magnetic collapse, and spin and orbital magnetic
moments at the Ir site vanish abruptly \cite{AUU18}.

Figure \ref{BND}(c) presents the {\it ab initio} energy band structure
of the $\beta$-Li$_2$IrO$_3$ for monoclinic $C2/c$ structure
calculated in fully relativistic Dirac GGA+SO approximation. The
calculations reveal that a pressure-induced structural phase
transition $Fddd$ $\rightarrow$ $C2/c$ at $P_c$ $\sim$4 GPa is
accompanied by the electronic phase transition from magnetic Mott
insulator to nonmagnetic (NM) band insulator. The crystal field at the
Ir site ($C1$ point symmetry) causes the splitting of 5$d$ orbitals
into five singlets $z^2$, $x^2-y^2$, $xy$, $yz$, and $xz$. The
electronic structure of the high-pressure $C2/c$ structure possesses
an empty peak in proximity to the Fermi level which almost coincides
with the position of the corresponding UHB $J_\text{eff}$ = 1/2 states
in Mott insulator at ambient pressure. However, these two peaks have
completely different nature. The formation of Ir$_2$ dimers in the
hyperhoneycomb lattice at high-pressure gives rise to bonding and
antibonding molecular-orbital states \cite{TKG+19}. In a new
coordinate system ($x'=(x+z)/\sqrt{2}$, $y'=(x-z)/\sqrt{2}$, $z'=y$)
with $d_{xz}$ orbital, directed along the dimer $Y$ bond the two
subbands with predominant $d_{xz}$ character can be seen in
Fig. \ref{BND}(c) at 0.7 eV above the Fermi level and at $-$1.7 eV
below the Fermi level. They can be assigned to the antibonding and the
bonding states of Ir$_2$ dimer molecules.  The large
bonding-antibonding splitting stabilizes the $d_{xz}$-orbital-dominant
anti-bonding state of $t_{2g}$ holes and makes the system a NM band
insulator. The remaining $d_{xy}$ and $d_{yz}$ orbitals are very
strongly mixed by SO coupling and cannot be separated from each other,
therefore we would called them $d_{xy}/d_{yz}$ orbitals. These
orbitals contribute to the subbands between the $d_{xz}$-bonding and
-antibonding subbands due to weaker hybridization between the
nearest-neighbor Ir atoms than that of $d_{xy}/d_{yz}$ orbitals. An
energy gap is formed between the mixed $d_{xy}/d_{yz}$ subbands and
the empty antibonding $d_{xz}$ subband. SO coupling separates the
$d_{xy}/d_{yz}$ bands into two groups $(d_{xy}/d_{yz})_1$ and
$(d_{xy}/d_{yz})_2$ situated at 0 to $-$0.3 eV and $-$0.5 to $-$1.4
eV, respectively.

Due to the strong distortion of IrO$_6$ octahedra in the $C2/c$
high-pressure phase the $e_g$ orbitals, which are almost degenerate at
ambient pressure, split to $z^2$ and $x^2-y^2$ states and become
broader occupying the 3$-$4.5 eV energy interval.

\section{I\lowercase{r} $L_3$ RIXS spectrum}
\label{sec:rixs}

\begin{figure}[tbp!]
\begin{center}
\includegraphics[width=0.99\columnwidth]{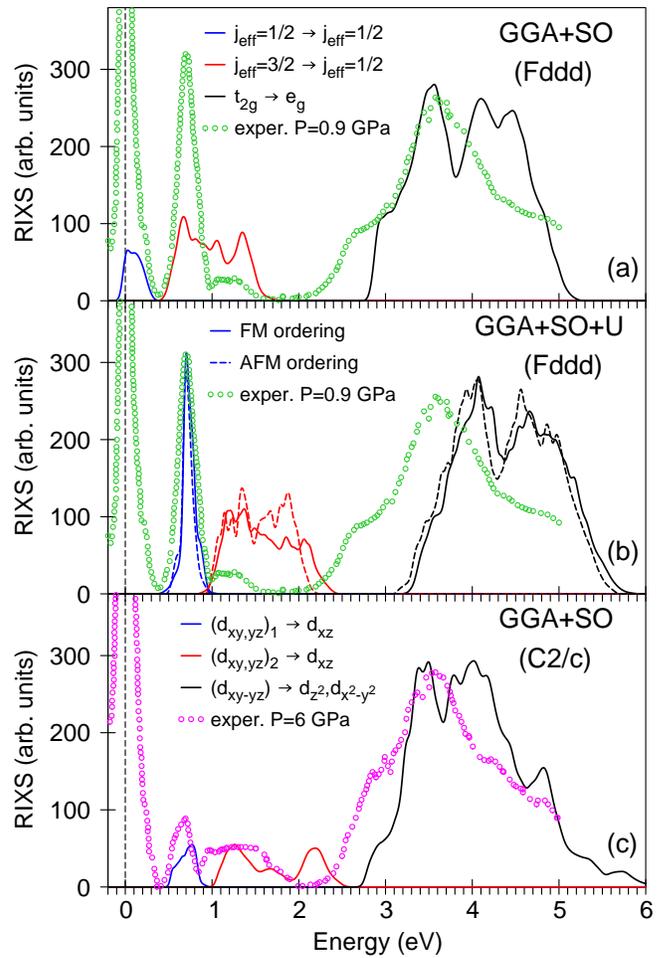}
\end{center}
\caption{\label{rixs}(Color online) (a) The experimental RIXS spectra
  at the Ir $L_3$ edge for the $\beta$-Li$_2$IrO$_3$ for the ambient
  pressure at room temperature \cite{TKG+19} (open green circles)
  compared with the theoretically calculated spectra in the GGA+SO
  approximation; (b) the experimental RIXS spectra at the Ir $L_3$
  edge for the $\beta$-Li$_2$IrO$_3$ for the ambient pressure
  \cite{TKG+19} (open green circles) compared with the theoretically
  calculated spectra in the GGA+SO+$U$ approach for the FM ordering
  along the $c$ direction (full curves) and for the AFM ordering along
  the $c$ direction (dashed curves); (c) The experimental RIXS spectra
  at the Ir $L_3$ edge for the $\beta$-Li$_2$IrO$_3$ for high-pressure
  phase ($C2/c$) above $P_c$ \cite{TKG+19} (open magenta circles)
  compared with the theoretically calculated spectra in the GGA+SO
  approximation.  }
\end{figure}

Figure \ref{rixs} presents the experimentally measured RIXS spectrum
at the Ir $L_3$ edge for the $\beta$-Li$_2$IrO$_3$ for the ambient
pressure \cite{TKG+19} (green open circles) compared with the
theoretically calculated ones in the GGA+SO (the upper panel) and
GGA+SO+$U$ approximations (middle panel) for the $Fddd$ phase. The
lower panel shows the experimental RIXS spectrum for high-pressure
phase ($C2/c$) above $P_c$ \cite{TKG+19} (open magenta circles)
compared with the theoretically calculated spectra in the GGA+SO
approximation.

The experimental RIXS spectrum at the Ir $L_3$ edge for the ambient
pressure in addition to the elastic scattering peak at 0 eV, possesses
a sharp narrow peak at around 0.7 eV, followed by a broad structure
between 1 eV and 2 eV, and a broad peak centered at around 3.5 eV. The
latter represents the excitations from Ir 5$d$ $t_{2g}$ to $e_g$
manifolds. The peak at $\sim$0.7 eV can be assigned to the local
excitation between the filled $J_\text{eff}$ = 1/2 and the empty
$J_\text{eff}$ = 1/2 states. The fine structure at 1 eV to 2 eV is
derived by the excitation between the filled $J_\text{eff}$ = 3/2 and
the empty $J_\text{eff}$ = 1/2 states.

The GGA+SO approximation fails to reproduce a sharp narrow peak at
around 0.7 eV. The corresponding peak is situated at 0.1 eV [blue
  curve in Fig. \ref{rixs}(a)] and has much smaller intensity in
comparison with the experimentally observed peak at 0.7 eV. On the
other hand, the theory correctly reproduces the energy position of the
wide peak at 3.5 eV responsible for the $t_{2g}$ $\rightarrow$ $e_g$
transitions [black curve in Fig. \ref{rixs}(a)]. In contrast, the
GGA+SO+$U$ approach well describes the shape and intensity of the LHB
$J_\text{eff}$ = 1/2 $\rightarrow$ UHB $J_\text{eff}$ = 1/2 peak,
however, it overestimates the intensity of $J_\text{eff}$ = 3/2
$\rightarrow$ $J_\text{eff}$ = 1/2 transitions presented by the red
curve in Fig. \ref{rixs}(b). In addition, the theoretically calculated
excitations from Ir 5$d$ $t_{2g}$ to $e_g$ manifolds are shifted
towards higher energies by approximately 0.7 eV in comparison with the
experiment. We calculated the RIXS spectra for the ferromagnetic (FM)
and antiferromegnetic (AFM) ordering along the $c$ direction and in
the $ab$ plane for the $Fddd$ phase and found that the type of
magnetic ordering weakly influences on the shape and intensity of
the major fine structures of Ir $L_3$ RIXS spectrum, especially low
energy peak at 0.7 eV [Fig. \ref{rixs}(b).

Due to applying of the Hubbard $U$, Ir 5$d$ occupied states are
shifted downward by $U_\text{eff}$/2 and the empty $d$ states are
shifted upward by this amount relative to the Fermi energy. The
Coulomb repulsion splits the half-filled $J_\text{eff}$ = 1/2 band
into an empty UHB and occupied LHB, open the energy gap and place the
UHB at 0.7 eV above the Fermi level in good agreement with the optical
\cite{AUU18} and RIXS measurements \cite{TKG+19}. Because we apply the
Hubbard $U$ to all the Ir 5$d$ states, it also shifts the empty
$t_{2g}$ states upward by $U_\text{eff}$/2 and places them at higher
energy than it detected in the RIXS experiment. It looks like the
correlation effects are mostly important for the 5$d$ states in close
vicinity of the Fermi level. The position of the $e_g$ states are
correctly reproduced by the GGA+SO approximation without taking into
account the electron correlations [see Fig. \ref{rixs}(a)].

Both the ambient pressure ($Fddd$) and the high-pressure ($C2/c$)
phases possess an empty peak in proximity to the Fermi level at around
0.7 eV (Fig. \ref{BND}), therefore one would expect similar RIXS
spectra at the low energy range below 1 eV. However, RIXS measurements
show the drastic reconstruction of the electronic structure associated
with the Ir dimerization \cite{TKG+19}. The 0.7 eV peak is suppressed
strongly in the high-pressure dimerized phase above 4 GPa. There are
some changes also at 1 to 2 eV energy interval. The wide peak at 3.5
eV associated with the excitations from Ir 5$d$ $t_{2g}$ to $e_g$
manifolds is broadened. The latter is due to the strong distortion of
the IrO$_6$ octahedra in the $C2/c$ high-pressure phase: the $e_g$
orbitals which are almost degenerated at ambient pressure split to
$z^2$ and $x^2-y^2$ states and became broader. The GGA+SO
approximation relatively well reproduces all the fine structures of
the RIXS spectrum under the pressure above the $P_c$ [see
  Fig. \ref{rixs}(c)]. The question is why the prominent peak at 0.7
eV is strongly suppressed in the RIXS spectra in the high-pressure
$C2/c$ phase.

The pressure-induced structural phase transition at $P_c$ $\sim$4 GPa
is accompanied by slight decrease of width of the $d_{xz}$ antibonding
empty states at 0.7 eV in comparison with the empty UHB
($J_\text{eff}$ = 1/2) states in the $Fddd$ phase (see Fig. 8 in
Ref. \cite{AUU18}). Therefore, one would expect more narrow RIXS peak
at 0.7 eV in the high-pressure phase and, at least, the similar
intensity in comparison with the $Fddd$ phase.

\begin{figure}[tbp!]
\begin{center}
\includegraphics[width=0.99\columnwidth]{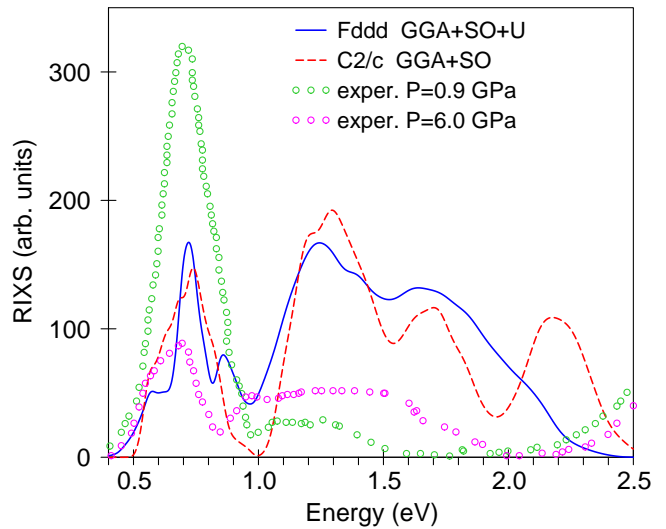}
\end{center}
\caption{\label{rixs_mme}(Color online) The comparison of RIXS spectra
  for the ambient (green circles) and high-pressure above $P_c$
  (magenta circles) phases\cite{TKG+19} with the constant matrix
  elements (JDOS) calculated for the ambient pressure phase ($Fddd$) in
  the GGA+SO+$U$ approximation (full blue curve) and for the
  high-pressure phase ($C2/c$) in the GGA+SO approach (dashed red
  curve). }
\end{figure}

The RIXS spectrum is a convolution of the densities of the occupied
and empty valence states [so called joint density of states (JDOS)],
weighted by appropriate matrix elements. Fig. \ref{rixs_mme} presents
comparison of the experimentally measured RIXS spectra for the ambient
and high-pressure phases \cite{TKG+19} with the JDOS calculated for
the ambient pressure phase ($Fddd$) in the GGA+SO+$U$ approximation
(the full blue curve) and for the high-pressure phase ($C2/c$) in the
GGA+SO approach (the dashed red curve). We see that without taking
into account the corresponding matrix elements the theoretically
calculated prominent peak at 0.7 eV has similar intensity for both the
phases and significantly differ from the experimentally measured
ones. Therefore, the reason of strong suppression of this peak in
high-pressure phase is the significantly different matrix elements in
these two phases.

\begin{figure}[tbp!]
\begin{center}
\includegraphics[width=0.99\columnwidth]{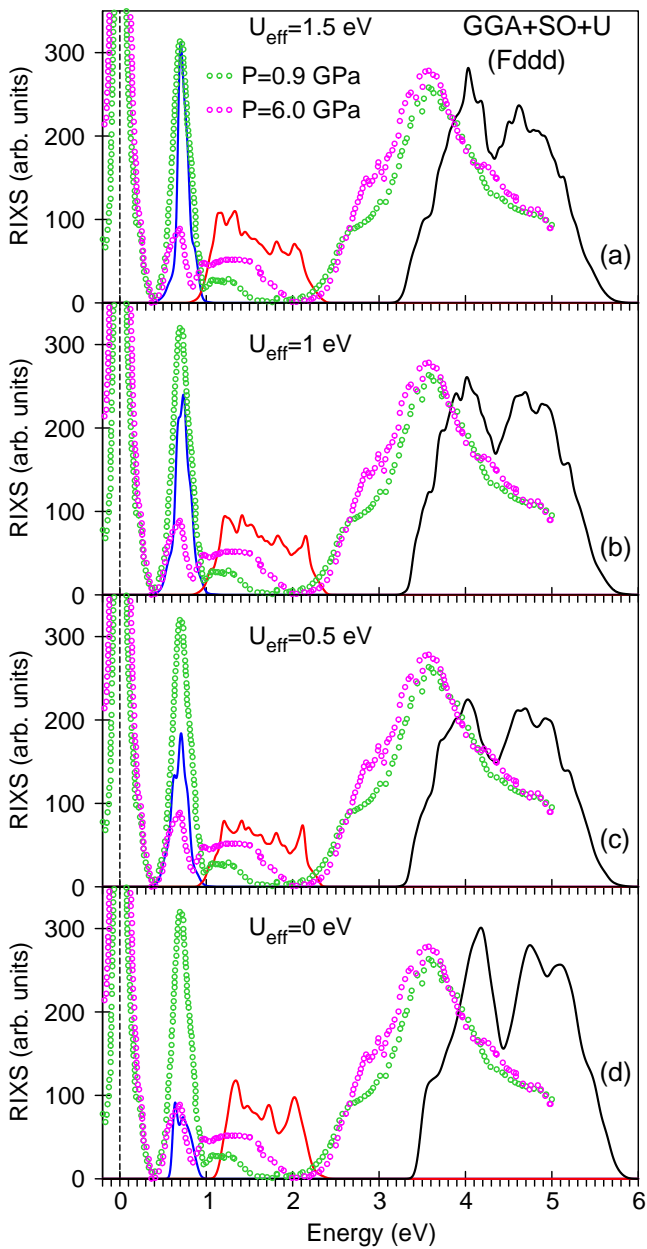}
\end{center}
\caption{\label{SO_rixs}(Color online) The experimental RIXS spectra
  at the Ir $L_3$ edge for the $\beta$-Li$_2$IrO$_3$ for the ambient
  pressure ($Fddd$) (open green circles) and high-pressure (open
  magenta circles) ($C2/c$) phases above $P_c$ \cite{TKG+19} compared
  with the theoretically calculated spectra in the GGA+SO+$U$
  approximation for different values of $U_{\text{eff}}$ at the Ir site. }
\end{figure}

There are two reasons why the matrix elements can be strongly altered
in the phase transition at the $P_c$. First, the prominent peak at 0.7
eV above the Fermi level is created by different orbitals ($t_{2g}$
states in $Fddd$ phase and almost pure $d_{xz}$ orbitals in
high-pressure $C2/c$ phase). Yang {\it at al.}  \cite{cm:YYA+09}
showed that angular matrix elements for dipole allowed transitions in
x-ray absorption at the $L_3$ edge strongly depend on the type of the
orbitals involved in the transitions. Second, the matrix elements can
be strongly modified due to different ground states of the phases: a
magnetic strongly correlated Mott insulator at ambient pressure and a
NM band insulator at $P_c$ with vanished Hubbard $U$.

At ambient pressure, $\beta$-Li$_2$IrO$_3$ shows a rather highly
symmetric honeycomb structure with a bond disproportionation of less
than 0.2\%. If pressure is increased from ambient pressure Ir-Ir bonds
monotonically shrink under the pressure and at the structural $Fddd
\rightarrow C2/c$ phase transition at $P_c$ = 4.4 GPa one of the $x/y$
bonds is strongly decreased to 2.6609 \AA\, \cite{VEG+17}, which is
even smaller than the Ir-Ir distance of 2.714 \AA\, in Ir metal. Such
the dimerization enhances the advantage of forming covalent bonds
which is accompanied by a magnetic collapse, spin and orbital magnetic
moments at the Ir site are vanished abruptly \cite{AUU18}. Pressure
reduces Coulomb correlation Hubbard parameter $U_{\text{eff}}$ which
decreases from 1.5 eV for ambient pressure to zero at $P_c$ = 4.4 GPa
\cite{AUU18} and reduces also the effective SO interaction at Ir
site. It is supported experimentally by the corresponding pressure
dependence of the branching ratio BR = $I_{L_3} / I_{L_2}$, where
$I_{L_{2,3}}$ is the integrated intensity of the isotropic x-ray
absorption spectra at the $L_{2,3}$ edges \cite{VEG+17}.

We examine the dependence of the RIXS spectra on the Hubbard $U$ at
the Ir site calculating the spectra for $U$=2.15 eV, 1.9 eV, 1.65 eV,
1.4 eV, 1.15 eV, 0.9 eV, and 0.65 eV. It corresponds to
$U_{\text{eff}}$=1.5 eV, 1.25 eV, 1.0 eV, 0.75 eV, 0,5 eV, 0.25 eV,
and 0 eV. The results of such calculations for the RIXS spectra
dependence on the $U_{\text{eff}}$ are shown in
Fig. \ref{SO_rixs}. The reduction of the $U_{\text{eff}}$ leads to
gradually reduction of the peak intensity of the prominent peak at 0.7
eV. For $U_{\text{eff}}$=0 eV this peak reduces to the value observed
in the high-pressure phase. We found that the energy gap is also
reduced by 0.102 eV, 0.202 eV and 0.301 eV for $U_{\text{eff}}$ = 1.25
eV, 1.0, and 0.75 eV, respectively. The gap is closed for
$U_{\text{eff}}$ = 0.5 eV. As a result, the prominent peak situated at
0.7 eV shifts towards smaller energy with decreasing of the
$U_{\text{eff}}$ parameter. To show more clearly how the peak
intensity at 0.7 eV depends on the $U_{\text{eff}}$ we keep this peak
artificially at 0.7 eV above the Fermi level in Fig. \ref{SO_rixs}.

We can conclude that the drastic reconstruction of the RIXS spectrum
associated with the structural $Fddd \rightarrow C2/c$ phase
transition at $P_c$ can be related to the Coulomb correlations
disappearing in the high-pressure $C2/c$ phase.

\section{Conclusions}

The electronic structure and RIXS spectra of the hyperhoneycomb
iridate $\beta$-Li$_2$IrO$_3$ were investigated theoretically within
the DFT-GGA approach in the framework of the fully relativistic
spin-polarized Dirac LMTO band-structure method, taking into account
Coulomb electron-electron correlations.

The GGA+SO approximation produces a metallic ground state in the
$\beta$-Li$_2$IrO$_3$, in contradiction with resistivity measurements
which claims that the $\beta$-Li$_2$IrO$_3$ is a spin-orbit Mott
insulator. To produce the correct ground state we have to take into
account Hubbard electron-electron correlations. The Coulomb repulsion
$U$ splits the half-filled $J_\text{eff}$ = 1/2 band into an empty
upper Hubbard band with pure $d_{5/2}$ character well separated from
the lower $J_\text{eff}$ = 1/2 Hubbard band and the $J_\text{eff}$ =
3/2 states below the Fermi level. We found that the GGA+SO+$U$
approach with Hubbard $U_\text{eff}$ = 1.5 eV well describes the RIXS
spectrum at the Ir $L_3$ edge for ambient pressure in the $Fddd$
phase.

The hyperhoneycomb iridate $\beta$-Li$_2$IrO$_3$, a spin-orbit Mott
insulator, is on the border of the magnetic ordering with relatively
weak Coulomb electron-electron correlations ($U_\text{eff}$ = 1.5
eV). The $\beta$-Li$_2$IrO$_3$ undergoes a pressure-induced structural
and magnetic phase transitions at $P_c$ $\sim$4 GPa with symmetry
lowering to the monoclinic $C2/c$. The structural phase transition is
accompanied by a dimerization of the previously equally long $x/y$
Ir-Ir bonds. We found remarkable NM ground states of the
$\beta$-Li$_2$IrO$_3$ at $P_c$, with a concomitant electronic phase
transition from a Mott insulator to band insulators.

The energy band structure of the $\beta$-Li$_2$IrO$_3$ for the ambient
pressure phase ($Fddd$) in the GGA+SO+$U$ approximation and for the
high-pressure phase ($C2/c$) possess similar peaks in the DOSs with
the same energy positions. However, these peaks have completely
different nature. There are empty UHB ($J_\text{eff}$ = 1/2) and
occupied LHB ($J_\text{eff}$ = 1/2) peaks at $\sim$0.7 eV and $-$0.2
eV, respectively, in the $Fddd$ phase. The similar peaks in
high-pressure $C2/c$ phase are assigned to the $d_{xz}$ antibonding
states of Ir$_2$ dimer molecules (at 0.7 eV) and mixture of the
($d_{xy}-d_{yz}$)$_1$ states. The peaks between $-$0.5 eV and $-$1.9
eV belong to the $J_\text{eff}$ = 3/2 states in $Fddd$ phase and the
($d_{xy}-d_{yz}$)$_2$ states plus the $d_{xz}$ molecular bonding
states in the high-pressure $C2/c$ phase.

The drastic reconstruction of the RIXS spectral peak at 0.7 eV
associated with the structural $Fddd \rightarrow C2/c$ phase
transition at $P_c$ can be related to disappearing of the Coulomb
correlations.

\section*{Acknowledgments}

We are thankful to Dr. Alexander Yaresko from Max-Planck-Institute FKF
in Stuttgart for very long and helpful discussions. V.N.A. gratefully
acknowledges the hospitality at the University of Bialystok during his
stay there.  D.A.K. gratefully acknowledges the hospitality at the
Max-Planck-Institute FKF during his stay in Stuttgart.

The studies were supported by the National Academy of Sciences of
Ukraine within the budget program KPKBK 6541230-3A "Support for the
development of priority areas of scientific research".


\begin{thebibliography}{44}
\expandafter\ifx\csname natexlab\endcsname\relax\def\natexlab#1{#1}\fi
\expandafter\ifx\csname bibnamefont\endcsname\relax
  \def\bibnamefont#1{#1}\fi
\expandafter\ifx\csname bibfnamefont\endcsname\relax
  \def\bibfnamefont#1{#1}\fi
\expandafter\ifx\csname citenamefont\endcsname\relax
  \def\citenamefont#1{#1}\fi
\expandafter\ifx\csname url\endcsname\relax
  \def\url#1{\texttt{#1}}\fi
\expandafter\ifx\csname urlprefix\endcsname\relax\def\urlprefix{URL }\fi
\providecommand{\bibinfo}[2]{#2}
\providecommand{\eprint}[2][]{\url{#2}}

\bibitem[{\citenamefont{Jackeli and Khaliullin}(2009)}]{JaKh09}
\bibinfo{author}{\bibfnamefont{G.}~\bibnamefont{Jackeli}} \bibnamefont{and}
  \bibinfo{author}{\bibfnamefont{G.}~\bibnamefont{Khaliullin}},
  \bibinfo{journal}{Phys. Rev. Lett.} \textbf{\bibinfo{volume}{102}},
  \bibinfo{pages}{017205} (\bibinfo{year}{2009}).

\bibitem[{\citenamefont{Witczak-Krempa
  et~al.}(2014)\citenamefont{Witczak-Krempa, Chen, Kim, and Balents}}]{WCK+14}
\bibinfo{author}{\bibfnamefont{W.}~\bibnamefont{Witczak-Krempa}},
  \bibinfo{author}{\bibfnamefont{G.}~\bibnamefont{Chen}},
  \bibinfo{author}{\bibfnamefont{Y.~B.} \bibnamefont{Kim}}, \bibnamefont{and}
  \bibinfo{author}{\bibfnamefont{L.}~\bibnamefont{Balents}},
  \bibinfo{journal}{Ann. Rev. Condens. Matter Phys.}
  \textbf{\bibinfo{volume}{89}}, \bibinfo{pages}{025003}
  (\bibinfo{year}{2014}).

\bibitem[{\citenamefont{Kimchi et~al.}(2014)\citenamefont{Kimchi, Analytis, and
  Vishwanath}}]{KAV14}
\bibinfo{author}{\bibfnamefont{I.}~\bibnamefont{Kimchi}},
  \bibinfo{author}{\bibfnamefont{J.~G.} \bibnamefont{Analytis}},
  \bibnamefont{and}
  \bibinfo{author}{\bibfnamefont{A.}~\bibnamefont{Vishwanath}},
  \bibinfo{journal}{Phys. Rev. B} \textbf{\bibinfo{volume}{90}},
  \bibinfo{pages}{205126} (\bibinfo{year}{2014}).

\bibitem[{\citenamefont{Kitaev}(2006)}]{Kit06}
\bibinfo{author}{\bibfnamefont{A.}~\bibnamefont{Kitaev}},
  \bibinfo{journal}{Ann. Phys.} \textbf{\bibinfo{volume}{321}},
  \bibinfo{pages}{2} (\bibinfo{year}{2006}).

\bibitem[{\citenamefont{Albrecht et~al.}(2016)\citenamefont{Albrecht,
  Higginbotham, Madsen, Kuemmeth, Jespersen, Nygrd, Krogstrup, and
  Marcus}}]{AHM+16}
\bibinfo{author}{\bibfnamefont{S.~M.} \bibnamefont{Albrecht}},
  \bibinfo{author}{\bibfnamefont{A.~P.} \bibnamefont{Higginbotham}},
  \bibinfo{author}{\bibfnamefont{M.}~\bibnamefont{Madsen}},
  \bibinfo{author}{\bibfnamefont{F.}~\bibnamefont{Kuemmeth}},
  \bibinfo{author}{\bibfnamefont{T.~S.} \bibnamefont{Jespersen}},
  \bibinfo{author}{\bibfnamefont{J.}~\bibnamefont{Nygrd}},
  \bibinfo{author}{\bibfnamefont{P.}~\bibnamefont{Krogstrup}},
  \bibnamefont{and} \bibinfo{author}{\bibfnamefont{C.~M.}
  \bibnamefont{Marcus}}, \bibinfo{journal}{Nature}
  \textbf{\bibinfo{volume}{531}}, \bibinfo{pages}{206} (\bibinfo{year}{2016}).

\bibitem[{\citenamefont{Winter et~al.}(2017)\citenamefont{Winter, Tsirlin,
  Daghofer, van~den Brink, Singh, Gegenwart, and Valenti}}]{WTD+17}
\bibinfo{author}{\bibfnamefont{S.~M.} \bibnamefont{Winter}},
  \bibinfo{author}{\bibfnamefont{A.~A.} \bibnamefont{Tsirlin}},
  \bibinfo{author}{\bibfnamefont{M.}~\bibnamefont{Daghofer}},
  \bibinfo{author}{\bibfnamefont{J.}~\bibnamefont{van~den Brink}},
  \bibinfo{author}{\bibfnamefont{Y.}~\bibnamefont{Singh}},
  \bibinfo{author}{\bibfnamefont{P.}~\bibnamefont{Gegenwart}},
  \bibnamefont{and} \bibinfo{author}{\bibfnamefont{R.}~\bibnamefont{Valenti}},
  \bibinfo{journal}{J. Phys.: Condens. Matter} \textbf{\bibinfo{volume}{29}},
  \bibinfo{pages}{493002} (\bibinfo{year}{2017}).

\bibitem[{\citenamefont{Schaffer et~al.}(2016)\citenamefont{Schaffer, Lee,
  Yang, and Kim}}]{SLY+16}
\bibinfo{author}{\bibfnamefont{R.}~\bibnamefont{Schaffer}},
  \bibinfo{author}{\bibfnamefont{E.~K.-H.} \bibnamefont{Lee}},
  \bibinfo{author}{\bibfnamefont{B.-J.} \bibnamefont{Yang}}, \bibnamefont{and}
  \bibinfo{author}{\bibfnamefont{Y.~B.} \bibnamefont{Kim}},
  \bibinfo{journal}{Rep. Prog. Phys.} \textbf{\bibinfo{volume}{79}},
  \bibinfo{pages}{094504} (\bibinfo{year}{2016}).

\bibitem[{\citenamefont{Ye et~al.}(2012)\citenamefont{Ye, Chi, Cao,
  Chakoumakos, Fernandez-Baca, Custelcean, Qi, Korneta, and Cao}}]{YCC+12}
\bibinfo{author}{\bibfnamefont{F.}~\bibnamefont{Ye}},
  \bibinfo{author}{\bibfnamefont{S.}~\bibnamefont{Chi}},
  \bibinfo{author}{\bibfnamefont{H.}~\bibnamefont{Cao}},
  \bibinfo{author}{\bibfnamefont{B.~C.} \bibnamefont{Chakoumakos}},
  \bibinfo{author}{\bibfnamefont{J.~A.} \bibnamefont{Fernandez-Baca}},
  \bibinfo{author}{\bibfnamefont{R.}~\bibnamefont{Custelcean}},
  \bibinfo{author}{\bibfnamefont{T.~F.} \bibnamefont{Qi}},
  \bibinfo{author}{\bibfnamefont{O.~B.} \bibnamefont{Korneta}},
  \bibnamefont{and} \bibinfo{author}{\bibfnamefont{G.}~\bibnamefont{Cao}},
  \bibinfo{journal}{Phys. Rev. B} \textbf{\bibinfo{volume}{85}},
  \bibinfo{pages}{180403} (\bibinfo{year}{2012}).

\bibitem[{\citenamefont{Shitade et~al.}(2009)\citenamefont{Shitade, Katsura,
  Kunes, Qi, Zhang, and Nagaosa}}]{SKK+09}
\bibinfo{author}{\bibfnamefont{A.}~\bibnamefont{Shitade}},
  \bibinfo{author}{\bibfnamefont{H.}~\bibnamefont{Katsura}},
  \bibinfo{author}{\bibfnamefont{J.}~\bibnamefont{Kunes}},
  \bibinfo{author}{\bibfnamefont{X.-L.} \bibnamefont{Qi}},
  \bibinfo{author}{\bibfnamefont{S.-C.} \bibnamefont{Zhang}}, \bibnamefont{and}
  \bibinfo{author}{\bibfnamefont{N.}~\bibnamefont{Nagaosa}},
  \bibinfo{journal}{Phys. Rev. Lett.} \textbf{\bibinfo{volume}{102}},
  \bibinfo{pages}{256403} (\bibinfo{year}{2009}).

\bibitem[{\citenamefont{Hasan and Kane}(2010)}]{HaKa10}
\bibinfo{author}{\bibfnamefont{M.~Z.} \bibnamefont{Hasan}} \bibnamefont{and}
  \bibinfo{author}{\bibfnamefont{C.~L.} \bibnamefont{Kane}},
  \bibinfo{journal}{Rev. Mod. Phys.} \textbf{\bibinfo{volume}{82}},
  \bibinfo{pages}{3045} (\bibinfo{year}{2010}).

\bibitem[{\citenamefont{Chaloupka et~al.}(2013)\citenamefont{Chaloupka,
  Jackeli, and Khaliullin}}]{CJK13}
\bibinfo{author}{\bibfnamefont{J.}~\bibnamefont{Chaloupka}},
  \bibinfo{author}{\bibfnamefont{G.}~\bibnamefont{Jackeli}}, \bibnamefont{and}
  \bibinfo{author}{\bibfnamefont{G.}~\bibnamefont{Khaliullin}},
  \bibinfo{journal}{Phys. Rev. Lett.} \textbf{\bibinfo{volume}{110}},
  \bibinfo{pages}{097204} (\bibinfo{year}{2013}).

\bibitem[{\citenamefont{Cao et~al.}(2013)\citenamefont{Cao, Qi, Li, Terzic,
  Cao, Yuan, Tovar, Murthy, and Kaul}}]{CQL+13}
\bibinfo{author}{\bibfnamefont{G.}~\bibnamefont{Cao}},
  \bibinfo{author}{\bibfnamefont{T.~F.} \bibnamefont{Qi}},
  \bibinfo{author}{\bibfnamefont{L.}~\bibnamefont{Li}},
  \bibinfo{author}{\bibfnamefont{J.}~\bibnamefont{Terzic}},
  \bibinfo{author}{\bibfnamefont{V.~S.} \bibnamefont{Cao}},
  \bibinfo{author}{\bibfnamefont{S.~J.} \bibnamefont{Yuan}},
  \bibinfo{author}{\bibfnamefont{M.}~\bibnamefont{Tovar}},
  \bibinfo{author}{\bibfnamefont{G.}~\bibnamefont{Murthy}}, \bibnamefont{and}
  \bibinfo{author}{\bibfnamefont{R.~K.} \bibnamefont{Kaul}},
  \bibinfo{journal}{Phys. Rev. B} \textbf{\bibinfo{volume}{88}},
  \bibinfo{pages}{220414} (\bibinfo{year}{2013}).

\bibitem[{\citenamefont{Biffin et~al.}(2014)\citenamefont{Biffin, Johnson,
  Choi, Freund, Manni, Bombardi, Manuel, Gegenwart, and Coldea}}]{BJC+14}
\bibinfo{author}{\bibfnamefont{A.}~\bibnamefont{Biffin}},
  \bibinfo{author}{\bibfnamefont{R.~D.} \bibnamefont{Johnson}},
  \bibinfo{author}{\bibfnamefont{S.}~\bibnamefont{Choi}},
  \bibinfo{author}{\bibfnamefont{F.}~\bibnamefont{Freund}},
  \bibinfo{author}{\bibfnamefont{S.}~\bibnamefont{Manni}},
  \bibinfo{author}{\bibfnamefont{A.}~\bibnamefont{Bombardi}},
  \bibinfo{author}{\bibfnamefont{P.}~\bibnamefont{Manuel}},
  \bibinfo{author}{\bibfnamefont{P.}~\bibnamefont{Gegenwart}},
  \bibnamefont{and} \bibinfo{author}{\bibfnamefont{R.}~\bibnamefont{Coldea}},
  \bibinfo{journal}{Phys. Rev. B} \textbf{\bibinfo{volume}{90}},
  \bibinfo{pages}{205116} (\bibinfo{year}{2014}).

\bibitem[{\citenamefont{Knolle et~al.}(2014)\citenamefont{Knolle, Chern,
  Kovrizhin, Moessner, and Perkins}}]{KCK+14}
\bibinfo{author}{\bibfnamefont{J.}~\bibnamefont{Knolle}},
  \bibinfo{author}{\bibfnamefont{G.-W.} \bibnamefont{Chern}},
  \bibinfo{author}{\bibfnamefont{D.~L.} \bibnamefont{Kovrizhin}},
  \bibinfo{author}{\bibfnamefont{R.}~\bibnamefont{Moessner}}, \bibnamefont{and}
  \bibinfo{author}{\bibfnamefont{N.~B.} \bibnamefont{Perkins}},
  \bibinfo{journal}{Phys. Rev. Lett.} \textbf{\bibinfo{volume}{113}},
  \bibinfo{pages}{187201} (\bibinfo{year}{2014}).

\bibitem[{\citenamefont{Takayama et~al.}(2015)\citenamefont{Takayama, Kato,
  Dinnebier, Nuss, Kono, Veiga, Fabbris, Haskel, and Takagi}}]{TKD+15}
\bibinfo{author}{\bibfnamefont{T.}~\bibnamefont{Takayama}},
  \bibinfo{author}{\bibfnamefont{A.}~\bibnamefont{Kato}},
  \bibinfo{author}{\bibfnamefont{R.}~\bibnamefont{Dinnebier}},
  \bibinfo{author}{\bibfnamefont{J.}~\bibnamefont{Nuss}},
  \bibinfo{author}{\bibfnamefont{H.}~\bibnamefont{Kono}},
  \bibinfo{author}{\bibfnamefont{L.~S.~I.} \bibnamefont{Veiga}},
  \bibinfo{author}{\bibfnamefont{G.}~\bibnamefont{Fabbris}},
  \bibinfo{author}{\bibfnamefont{D.}~\bibnamefont{Haskel}}, \bibnamefont{and}
  \bibinfo{author}{\bibfnamefont{H.}~\bibnamefont{Takagi}},
  \bibinfo{journal}{Phys. Rev. Lett.} \textbf{\bibinfo{volume}{114}},
  \bibinfo{pages}{077202} (\bibinfo{year}{2015}).

\bibitem[{\citenamefont{Schaffer et~al.}(2015)\citenamefont{Schaffer, Lee, Lu,
  and Kim}}]{SLLK15}
\bibinfo{author}{\bibfnamefont{R.}~\bibnamefont{Schaffer}},
  \bibinfo{author}{\bibfnamefont{E.~K.-H.} \bibnamefont{Lee}},
  \bibinfo{author}{\bibfnamefont{Y.-M.} \bibnamefont{Lu}}, \bibnamefont{and}
  \bibinfo{author}{\bibfnamefont{Y.~B.} \bibnamefont{Kim}},
  \bibinfo{journal}{Phys. Rev. Lett.} \textbf{\bibinfo{volume}{114}},
  \bibinfo{pages}{116803} (\bibinfo{year}{2015}).

\bibitem[{\citenamefont{Glamazda et~al.}(2016)\citenamefont{Glamazda, Lemmens,
  Do, Choi, and Choi}}]{GLD+16}
\bibinfo{author}{\bibfnamefont{A.}~\bibnamefont{Glamazda}},
  \bibinfo{author}{\bibfnamefont{P.}~\bibnamefont{Lemmens}},
  \bibinfo{author}{\bibfnamefont{S.~H.} \bibnamefont{Do}},
  \bibinfo{author}{\bibfnamefont{Y.~S.} \bibnamefont{Choi}}, \bibnamefont{and}
  \bibinfo{author}{\bibfnamefont{K.~Y.} \bibnamefont{Choi}},
  \bibinfo{journal}{Nat. Commun.} \textbf{\bibinfo{volume}{7}},
  \bibinfo{pages}{12286} (\bibinfo{year}{2016}).

\bibitem[{\citenamefont{Kim et~al.}(2008)\citenamefont{Kim, Jin, Moon, Kim,
  Park, Leem, Yu, Noh, Kim, Oh et~al.}}]{KJM+08}
\bibinfo{author}{\bibfnamefont{B.~J.} \bibnamefont{Kim}},
  \bibinfo{author}{\bibfnamefont{H.}~\bibnamefont{Jin}},
  \bibinfo{author}{\bibfnamefont{S.~J.} \bibnamefont{Moon}},
  \bibinfo{author}{\bibfnamefont{J.-Y.} \bibnamefont{Kim}},
  \bibinfo{author}{\bibfnamefont{B.-G.} \bibnamefont{Park}},
  \bibinfo{author}{\bibfnamefont{C.~S.} \bibnamefont{Leem}},
  \bibinfo{author}{\bibfnamefont{J.}~\bibnamefont{Yu}},
  \bibinfo{author}{\bibfnamefont{T.~W.} \bibnamefont{Noh}},
  \bibinfo{author}{\bibfnamefont{C.}~\bibnamefont{Kim}},
  \bibinfo{author}{\bibfnamefont{S.-J.} \bibnamefont{Oh}},
  \bibnamefont{et~al.}, \bibinfo{journal}{Phys. Rev. Lett.}
  \textbf{\bibinfo{volume}{101}}, \bibinfo{pages}{076402}
  (\bibinfo{year}{2008}).

\bibitem[{\citenamefont{Kim et~al.}(2016)\citenamefont{Kim, Kim, and
  Kee}}]{KKK16}
\bibinfo{author}{\bibfnamefont{H.-S.} \bibnamefont{Kim}},
  \bibinfo{author}{\bibfnamefont{Y.~B.} \bibnamefont{Kim}}, \bibnamefont{and}
  \bibinfo{author}{\bibfnamefont{H.-Y.} \bibnamefont{Kee}},
  \bibinfo{journal}{Phys. Rev. B} \textbf{\bibinfo{volume}{94}},
  \bibinfo{pages}{245127} (\bibinfo{year}{2016}).

\bibitem[{\citenamefont{Lee and Kim}(2015)}]{LeKi15}
\bibinfo{author}{\bibfnamefont{E.~K.-H.} \bibnamefont{Lee}} \bibnamefont{and}
  \bibinfo{author}{\bibfnamefont{Y.~B.} \bibnamefont{Kim}},
  \bibinfo{journal}{Phys. Rev. B} \textbf{\bibinfo{volume}{91}},
  \bibinfo{pages}{064407} (\bibinfo{year}{2015}).

\bibitem[{\citenamefont{Kimchi et~al.}(2015)\citenamefont{Kimchi, Coldea, and
  Vishwanath}}]{KCV15}
\bibinfo{author}{\bibfnamefont{I.}~\bibnamefont{Kimchi}},
  \bibinfo{author}{\bibfnamefont{R.}~\bibnamefont{Coldea}}, \bibnamefont{and}
  \bibinfo{author}{\bibfnamefont{A.}~\bibnamefont{Vishwanath}},
  \bibinfo{journal}{Phys. Rev. B} \textbf{\bibinfo{volume}{91}},
  \bibinfo{pages}{245134} (\bibinfo{year}{2015}).

\bibitem[{\citenamefont{Katukuri et~al.}(2016)\citenamefont{Katukuri, Yadav,
  Hozoi, Nishimoto, and van~den Brink}}]{KYH+16}
\bibinfo{author}{\bibfnamefont{V.~M.} \bibnamefont{Katukuri}},
  \bibinfo{author}{\bibfnamefont{R.}~\bibnamefont{Yadav}},
  \bibinfo{author}{\bibfnamefont{L.}~\bibnamefont{Hozoi}},
  \bibinfo{author}{\bibfnamefont{S.}~\bibnamefont{Nishimoto}},
  \bibnamefont{and} \bibinfo{author}{\bibfnamefont{J.}~\bibnamefont{van~den
  Brink}}, \bibinfo{journal}{Sci. Rep.} \textbf{\bibinfo{volume}{6}},
  \bibinfo{pages}{29585} (\bibinfo{year}{2016}).

\bibitem[{\citenamefont{Ruiz et~al.}(2017)\citenamefont{Ruiz, Frano, Breznay,
  Kimchi, Helm, Oswald, Chan, Birgeneau, Islam, and Analytis}}]{RFB+17}
\bibinfo{author}{\bibfnamefont{A.}~\bibnamefont{Ruiz}},
  \bibinfo{author}{\bibfnamefont{A.}~\bibnamefont{Frano}},
  \bibinfo{author}{\bibfnamefont{N.~P.} \bibnamefont{Breznay}},
  \bibinfo{author}{\bibfnamefont{I.}~\bibnamefont{Kimchi}},
  \bibinfo{author}{\bibfnamefont{T.}~\bibnamefont{Helm}},
  \bibinfo{author}{\bibfnamefont{I.}~\bibnamefont{Oswald}},
  \bibinfo{author}{\bibfnamefont{J.~Y.} \bibnamefont{Chan}},
  \bibinfo{author}{\bibfnamefont{R.}~\bibnamefont{Birgeneau}},
  \bibinfo{author}{\bibfnamefont{Z.}~\bibnamefont{Islam}}, \bibnamefont{and}
  \bibinfo{author}{\bibfnamefont{J.~G.} \bibnamefont{Analytis}},
  \bibinfo{journal}{Nature Comm.} \textbf{\bibinfo{volume}{8}},
  \bibinfo{pages}{961} (\bibinfo{year}{2017}).

\bibitem[{\citenamefont{Ducatman et~al.}(2018)\citenamefont{Ducatman,
  Rousochatzakis, and Perkins}}]{DRP18}
\bibinfo{author}{\bibfnamefont{S.}~\bibnamefont{Ducatman}},
  \bibinfo{author}{\bibfnamefont{I.}~\bibnamefont{Rousochatzakis}},
  \bibnamefont{and} \bibinfo{author}{\bibfnamefont{N.~B.}
  \bibnamefont{Perkins}}, \bibinfo{journal}{Phys. Rev. B}
  \textbf{\bibinfo{volume}{97}}, \bibinfo{pages}{125125}
  (\bibinfo{year}{2018}).

\bibitem[{\citenamefont{Majumder et~al.}(2018)\citenamefont{Majumder, Manna,
  Simutis, Orain, Dey, Freund, Jesche, Khasanov, Biswas, Bykova
  et~al.}}]{MMS+18}
\bibinfo{author}{\bibfnamefont{M.}~\bibnamefont{Majumder}},
  \bibinfo{author}{\bibfnamefont{R.~S.} \bibnamefont{Manna}},
  \bibinfo{author}{\bibfnamefont{G.}~\bibnamefont{Simutis}},
  \bibinfo{author}{\bibfnamefont{J.~C.} \bibnamefont{Orain}},
  \bibinfo{author}{\bibfnamefont{T.}~\bibnamefont{Dey}},
  \bibinfo{author}{\bibfnamefont{F.}~\bibnamefont{Freund}},
  \bibinfo{author}{\bibfnamefont{A.}~\bibnamefont{Jesche}},
  \bibinfo{author}{\bibfnamefont{R.}~\bibnamefont{Khasanov}},
  \bibinfo{author}{\bibfnamefont{P.~K.} \bibnamefont{Biswas}},
  \bibinfo{author}{\bibfnamefont{E.}~\bibnamefont{Bykova}},
  \bibnamefont{et~al.}, \bibinfo{journal}{Phys. Rev. Lett.}
  \textbf{\bibinfo{volume}{120}}, \bibinfo{pages}{237202}
  (\bibinfo{year}{2018}).

\bibitem[{\citenamefont{Hermann et~al.}(2018)\citenamefont{Hermann, Altmeyer,
  Ebad-Allah, Freund, Jesche, Tsirlin, Hanfland, Gegenwart, Mazin, Khomskii
  et~al.}}]{HAE+18}
\bibinfo{author}{\bibfnamefont{V.}~\bibnamefont{Hermann}},
  \bibinfo{author}{\bibfnamefont{M.}~\bibnamefont{Altmeyer}},
  \bibinfo{author}{\bibfnamefont{J.}~\bibnamefont{Ebad-Allah}},
  \bibinfo{author}{\bibfnamefont{F.}~\bibnamefont{Freund}},
  \bibinfo{author}{\bibfnamefont{A.}~\bibnamefont{Jesche}},
  \bibinfo{author}{\bibfnamefont{A.~A.} \bibnamefont{Tsirlin}},
  \bibinfo{author}{\bibfnamefont{M.}~\bibnamefont{Hanfland}},
  \bibinfo{author}{\bibfnamefont{P.}~\bibnamefont{Gegenwart}},
  \bibinfo{author}{\bibfnamefont{I.~I.} \bibnamefont{Mazin}},
  \bibinfo{author}{\bibfnamefont{D.~I.} \bibnamefont{Khomskii}},
  \bibnamefont{et~al.}, \bibinfo{journal}{Phys. Rev. B}
  \textbf{\bibinfo{volume}{97}}, \bibinfo{pages}{020104(R)}
  (\bibinfo{year}{2018}).

\bibitem[{\citenamefont{Ament et~al.}(2011)\citenamefont{Ament, van Veenendaal,
  Devereaux, Hill, and van~den Brink}}]{AVD+11}
\bibinfo{author}{\bibfnamefont{L.~J.~P.} \bibnamefont{Ament}},
  \bibinfo{author}{\bibfnamefont{M.}~\bibnamefont{van Veenendaal}},
  \bibinfo{author}{\bibfnamefont{T.~P.} \bibnamefont{Devereaux}},
  \bibinfo{author}{\bibfnamefont{J.~P.} \bibnamefont{Hill}}, \bibnamefont{and}
  \bibinfo{author}{\bibfnamefont{J.}~\bibnamefont{van~den Brink}},
  \bibinfo{journal}{Rev. Mod. Phys.} \textbf{\bibinfo{volume}{83}},
  \bibinfo{pages}{705} (\bibinfo{year}{2011}).

\bibitem[{\citenamefont{Takayama et~al.}(2019)\citenamefont{Takayama,
  Krajewska, Gibbs, Yaresko, Ishii, Yamaoka, Ishii, Hiraoka, Funnell, Bull
  et~al.}}]{TKG+19}
\bibinfo{author}{\bibfnamefont{T.}~\bibnamefont{Takayama}},
  \bibinfo{author}{\bibfnamefont{A.}~\bibnamefont{Krajewska}},
  \bibinfo{author}{\bibfnamefont{A.~S.} \bibnamefont{Gibbs}},
  \bibinfo{author}{\bibfnamefont{A.~N.} \bibnamefont{Yaresko}},
  \bibinfo{author}{\bibfnamefont{H.}~\bibnamefont{Ishii}},
  \bibinfo{author}{\bibfnamefont{H.}~\bibnamefont{Yamaoka}},
  \bibinfo{author}{\bibfnamefont{K.}~\bibnamefont{Ishii}},
  \bibinfo{author}{\bibfnamefont{N.}~\bibnamefont{Hiraoka}},
  \bibinfo{author}{\bibfnamefont{N.~P.} \bibnamefont{Funnell}},
  \bibinfo{author}{\bibfnamefont{C.~L.} \bibnamefont{Bull}},
  \bibnamefont{et~al.}, \bibinfo{journal}{Phys. Rev. B}
  \textbf{\bibinfo{volume}{99}}, \bibinfo{pages}{125127}
  (\bibinfo{year}{2019}).

\bibitem[{\citenamefont{Antonov et~al.}(2018)\citenamefont{Antonov, Uba, and
  Uba}}]{AUU18}
\bibinfo{author}{\bibfnamefont{V.~N.} \bibnamefont{Antonov}},
  \bibinfo{author}{\bibfnamefont{S.}~\bibnamefont{Uba}}, \bibnamefont{and}
  \bibinfo{author}{\bibfnamefont{L.}~\bibnamefont{Uba}},
  \bibinfo{journal}{Phys. Rev. B} \textbf{\bibinfo{volume}{98}},
  \bibinfo{pages}{245113} (\bibinfo{year}{2018}).

\bibitem[{\citenamefont{Veiga et~al.}(2017)\citenamefont{Veiga, Etter,
  Glazyrin, Sun, C.~A.~Escanhoela, Fabbris, Mardegan, Malavi, Deng,
  Stavropoulos et~al.}}]{VEG+17}
\bibinfo{author}{\bibfnamefont{L.~S.~I.} \bibnamefont{Veiga}},
  \bibinfo{author}{\bibfnamefont{M.}~\bibnamefont{Etter}},
  \bibinfo{author}{\bibfnamefont{K.}~\bibnamefont{Glazyrin}},
  \bibinfo{author}{\bibfnamefont{F.}~\bibnamefont{Sun}},
  \bibinfo{author}{\bibfnamefont{J.}~\bibnamefont{C.~A.~Escanhoela}},
  \bibinfo{author}{\bibfnamefont{G.}~\bibnamefont{Fabbris}},
  \bibinfo{author}{\bibfnamefont{J.~R.~L.} \bibnamefont{Mardegan}},
  \bibinfo{author}{\bibfnamefont{P.~S.} \bibnamefont{Malavi}},
  \bibinfo{author}{\bibfnamefont{Y.}~\bibnamefont{Deng}},
  \bibinfo{author}{\bibfnamefont{P.~P.} \bibnamefont{Stavropoulos}},
  \bibnamefont{et~al.}, \bibinfo{journal}{Phys. Rev. B}
  \textbf{\bibinfo{volume}{96}}, \bibinfo{pages}{140402(R)}
  (\bibinfo{year}{2017}).

\bibitem[{\citenamefont{Nemoshkalenko et~al.}(1983)\citenamefont{Nemoshkalenko,
  Krasovskii, Antonov, Antonov, Fleck, Wonn, and Ziesche}}]{NKA+83}
\bibinfo{author}{\bibfnamefont{V.~V.} \bibnamefont{Nemoshkalenko}},
  \bibinfo{author}{\bibfnamefont{A.~E.} \bibnamefont{Krasovskii}},
  \bibinfo{author}{\bibfnamefont{V.~N.} \bibnamefont{Antonov}},
  \bibinfo{author}{\bibfnamefont{V.~N.} \bibnamefont{Antonov}},
  \bibinfo{author}{\bibfnamefont{U.}~\bibnamefont{Fleck}},
  \bibinfo{author}{\bibfnamefont{H.}~\bibnamefont{Wonn}}, \bibnamefont{and}
  \bibinfo{author}{\bibfnamefont{P.}~\bibnamefont{Ziesche}},
  \bibinfo{journal}{Phys. status solidi B} \textbf{\bibinfo{volume}{120}},
  \bibinfo{pages}{283} (\bibinfo{year}{1983}).

\bibitem[{\citenamefont{Arola et~al.}(1997)\citenamefont{Arola, Strange, and
  Gyorffy}}]{ASG97}
\bibinfo{author}{\bibfnamefont{E.}~\bibnamefont{Arola}},
  \bibinfo{author}{\bibfnamefont{P.}~\bibnamefont{Strange}}, \bibnamefont{and}
  \bibinfo{author}{\bibfnamefont{B.~L.} \bibnamefont{Gyorffy}},
  \bibinfo{journal}{Phys. Rev. B} \textbf{\bibinfo{volume}{55}},
  \bibinfo{pages}{472} (\bibinfo{year}{1997}).

\bibitem[{\citenamefont{Lehmann and Taut}(1972)}]{LeTa72}
\bibinfo{author}{\bibfnamefont{G.}~\bibnamefont{Lehmann}} \bibnamefont{and}
  \bibinfo{author}{\bibfnamefont{M.}~\bibnamefont{Taut}},
  \bibinfo{journal}{Phys. status solidi B} \textbf{\bibinfo{volume}{54}},
  \bibinfo{pages}{469} (\bibinfo{year}{1972}).

\bibitem[{\citenamefont{Kukusta and Yaresko}(2018,
  unpublished)}]{unpub:KuYar18}
\bibinfo{author}{\bibfnamefont{D.~A.} \bibnamefont{Kukusta}} \bibnamefont{and}
  \bibinfo{author}{\bibfnamefont{A.~N.} \bibnamefont{Yaresko}}, in
  \emph{\bibinfo{booktitle}{Resonant inelastic x-ray scattering spectra from
  band structure calculations}} (\bibinfo{address}{Workshop "Strongly
  Correlated Electron Systems", Ringberg Castle, Germany}, \bibinfo{year}{2018,
  unpublished}).

\bibitem[{\citenamefont{Antonov et~al.}(2006)\citenamefont{Antonov, Jepsen,
  Yaresko, and Shpak}}]{AJY+06}
\bibinfo{author}{\bibfnamefont{V.~N.} \bibnamefont{Antonov}},
  \bibinfo{author}{\bibfnamefont{O.}~\bibnamefont{Jepsen}},
  \bibinfo{author}{\bibfnamefont{A.~N.} \bibnamefont{Yaresko}},
  \bibnamefont{and} \bibinfo{author}{\bibfnamefont{A.~P.} \bibnamefont{Shpak}},
  \bibinfo{journal}{J. Appl. Phys.} \textbf{\bibinfo{volume}{100}},
  \bibinfo{pages}{043711} (\bibinfo{year}{2006}).

\bibitem[{\citenamefont{Antonov et~al.}(2007)\citenamefont{Antonov, Harmon,
  Yaresko, and Shpak}}]{AHY+07b}
\bibinfo{author}{\bibfnamefont{V.~N.} \bibnamefont{Antonov}},
  \bibinfo{author}{\bibfnamefont{B.~N.} \bibnamefont{Harmon}},
  \bibinfo{author}{\bibfnamefont{A.~N.} \bibnamefont{Yaresko}},
  \bibnamefont{and} \bibinfo{author}{\bibfnamefont{A.~P.} \bibnamefont{Shpak}},
  \bibinfo{journal}{Phys. Rev. B} \textbf{\bibinfo{volume}{75}},
  \bibinfo{pages}{184422} (\bibinfo{year}{2007}).

\bibitem[{\citenamefont{Antonov et~al.}(2010)\citenamefont{Antonov, Yaresko,
  and Jepsen}}]{AYJ10}
\bibinfo{author}{\bibfnamefont{V.~N.} \bibnamefont{Antonov}},
  \bibinfo{author}{\bibfnamefont{A.~N.} \bibnamefont{Yaresko}},
  \bibnamefont{and} \bibinfo{author}{\bibfnamefont{O.}~\bibnamefont{Jepsen}},
  \bibinfo{journal}{Phys. Rev. B} \textbf{\bibinfo{volume}{81}},
  \bibinfo{pages}{075209} (\bibinfo{year}{2010}).

\bibitem[{\citenamefont{Antonov et~al.}(2020)\citenamefont{Antonov, Kukusta,
  and Bekenov}}]{AKB20}
\bibinfo{author}{\bibfnamefont{V.~N.} \bibnamefont{Antonov}},
  \bibinfo{author}{\bibfnamefont{D.~A.} \bibnamefont{Kukusta}},
  \bibnamefont{and} \bibinfo{author}{\bibfnamefont{L.~V.}
  \bibnamefont{Bekenov}}, \bibinfo{journal}{Phys. Rev. B}
  \textbf{\bibinfo{volume}{102}}, \bibinfo{pages}{195134}
  (\bibinfo{year}{2020}).

\bibitem[{\citenamefont{Andersen}(1975)}]{And75}
\bibinfo{author}{\bibfnamefont{O.~K.} \bibnamefont{Andersen}},
  \bibinfo{journal}{Phys. Rev. B} \textbf{\bibinfo{volume}{12}},
  \bibinfo{pages}{3060} (\bibinfo{year}{1975}).

\bibitem[{\citenamefont{Perlov et~al.}(1995, unpublished)\citenamefont{Perlov,
  Yaresko, and Antonov}}]{PYLMTO}
\bibinfo{author}{\bibfnamefont{A.~Y.} \bibnamefont{Perlov}},
  \bibinfo{author}{\bibfnamefont{A.~N.} \bibnamefont{Yaresko}},
  \bibnamefont{and} \bibinfo{author}{\bibfnamefont{V.~N.}
  \bibnamefont{Antonov}}, \bibinfo{journal}{{PY-LMTO}, A Spin-polarized
  Relativistic {L}inear {M}uffin-tin {O}rbitals Package for Electronic
  Structure Calculations}  (\bibinfo{year}{1995, unpublished}).

\bibitem[{\citenamefont{Perdew et~al.}(1996)\citenamefont{Perdew, Burke, and
  Ernzerhof}}]{PBE96}
\bibinfo{author}{\bibfnamefont{J.~P.} \bibnamefont{Perdew}},
  \bibinfo{author}{\bibfnamefont{K.}~\bibnamefont{Burke}}, \bibnamefont{and}
  \bibinfo{author}{\bibfnamefont{M.}~\bibnamefont{Ernzerhof}},
  \bibinfo{journal}{Phys. Rev. Lett.} \textbf{\bibinfo{volume}{77}},
  \bibinfo{pages}{3865} (\bibinfo{year}{1996}).

\bibitem[{\citenamefont{Bl\"ochl et~al.}(1994)\citenamefont{Bl\"ochl, Jepsen,
  and Andersen}}]{BJA94}
\bibinfo{author}{\bibfnamefont{P.~E.} \bibnamefont{Bl\"ochl}},
  \bibinfo{author}{\bibfnamefont{O.}~\bibnamefont{Jepsen}}, \bibnamefont{and}
  \bibinfo{author}{\bibfnamefont{O.~K.} \bibnamefont{Andersen}},
  \bibinfo{journal}{Phys. Rev. B} \textbf{\bibinfo{volume}{49}},
  \bibinfo{pages}{16223} (\bibinfo{year}{1994}).

\bibitem[{\citenamefont{Yaresko et~al.}(2003)\citenamefont{Yaresko, Antonov,
  and Fulde}}]{YAF03}
\bibinfo{author}{\bibfnamefont{A.~N.} \bibnamefont{Yaresko}},
  \bibinfo{author}{\bibfnamefont{V.~N.} \bibnamefont{Antonov}},
  \bibnamefont{and} \bibinfo{author}{\bibfnamefont{P.}~\bibnamefont{Fulde}},
  \bibinfo{journal}{Phys. Rev. B} \textbf{\bibinfo{volume}{67}},
  \bibinfo{pages}{155103} (\bibinfo{year}{2003}).

\bibitem[{\citenamefont{Yang et~al.}(2009)\citenamefont{Yang, Yaresko, Antonov,
  and Andersen}}]{cm:YYA+09}
\bibinfo{author}{\bibfnamefont{X.}~\bibnamefont{Yang}},
  \bibinfo{author}{\bibfnamefont{A.~N.} \bibnamefont{Yaresko}},
  \bibinfo{author}{\bibfnamefont{V.~N.} \bibnamefont{Antonov}},
  \bibnamefont{and} \bibinfo{author}{\bibfnamefont{O.~K.}
  \bibnamefont{Andersen}}, \bibinfo{journal}{preprint cond-mat/0911.4349v}
  (\bibinfo{year}{2009}).

\end{thebibliography}

\newcommand{\noopsort}[1]{} \newcommand{\printfirst}[2]{#1}
  \newcommand{\singleletter}[1]{#1} \newcommand{\switchargs}[2]{#2#1}

\end{document}